\documentclass[showpacs,preprintnumbers,amsmath,prl,graphicx,amssymb,superscriptaddress, twocolumn ]{revtex4}
\usepackage{graphicx}
\usepackage{dcolumn}
\usepackage{bm}
\usepackage{psfrag}
\usepackage{txfonts}
\newcommand{\snoccfluxunc}{1.59^{+0.08}_{-0.07}\mbox{(stat)}^{+0.06}_{-0.08}\mbox{(syst)}} 
\newcommand{\snoesfluxunc}{2.21^{+0.31}_{-0.26}\mbox{(stat)}~\pm{0.10}~\mbox{(syst)}} 
\newcommand{\snoncfluxunc}{5.21\pm 0.27~\mbox{(stat)}~\pm0.38~\mbox{(syst)}} 
\newcommand{\snoccfluxconc}{1.70\pm0.07\mbox{(stat)}^{+0.09}_{-0.10}\mbox{(syst)}} 
\newcommand{\snoesfluxconc}{2.13^{+0.29}_{-0.28}\mbox{(stat)}^{+0.15}_{-0.08}\mbox{(syst)}} 
\newcommand{\snoncfluxconc}{4.90\pm 0.24~\mbox{(stat)}^{+0.29}_{-0.27}\mbox{(syst)}} 
\newcommand{\snoccncratio}{0.306\pm0.026~\mbox{(stat)}~\pm0.024~\mbox{(syst)}}
\newcommand{\nccfit}{1339.6^{+63.8}_{-61.5}} 
\newcommand{\nesfit}{170.3^{+23.9}_{-20.1}} 
\newcommand{\nncfit}{1344.2^{+69.8}_{-69.0}}  
\newcommand{\nncbefit}{84.5^{+34.5}_{-33.6}}
\newcommand{\nsigmamax}{5.4}

\renewcommand{\today}{\number\day\space\ifcase\month\or January\or 
 February\or March\or April\or May\or June\or July\or August\or 
 September\or October\or November\or December\fi\space\number\year}

\begin{document}
\title{Measurement of the Total Active $^{\bm{8}}$B Solar Neutrino Flux at the Sudbury Neutrino Observatory with Enhanced Neutral Current Sensitivity}
\newcommand{\ubc}{Department of Physics and Astronomy, University of 
British Columbia, Vancouver, BC V6T 1Z1 Canada}
\newcommand{\bnl}{Chemistry Department, Brookhaven National 
Laboratory,  Upton, NY 11973-5000}
\newcommand{\carleton}{Ottawa-Carleton Institute for Physics, Department of Physics, Carleton University, Ottawa, Ontario K1S 5B6 Canada}
\newcommand{\uog}{Physics Department, University of Guelph,  
Guelph, Ontario N1G 2W1 Canada}
\newcommand{\lu}{Department of Physics and Astronomy, Laurentian 
University, Sudbury, Ontario P3E 2C6 Canada}
\newcommand{\lbnl}{Institute for Nuclear and Particle Astrophysics and 
Nuclear Science Division, Lawrence Berkeley National Laboratory, Berkeley, CA 94720}
\newcommand{\lanl}{Los Alamos National Laboratory, Los Alamos, NM 87545}
\newcommand{\oxford}{Department of Physics, University of Oxford, 
Denys Wilkinson Building, Keble Road, Oxford, OX1 3RH, UK}
\newcommand{\penn}{Department of Physics and Astronomy, University of 
Pennsylvania, Philadelphia, PA 19104-6396}
\newcommand{\queens}{Department of Physics, Queen's University, 
Kingston, Ontario K7L 3N6 Canada}
\newcommand{\uw}{Center for Experimental Nuclear Physics and Astrophysics, 
and Department of Physics, University of Washington, Seattle, WA 98195}
\newcommand{\triumf}{TRIUMF, 4004 Wesbrook Mall, Vancouver, BC V6T 2A3, Canada}
\newcommand{\ralsuss}{Rutherford Appleton Laboratory, Chilton, Didcot, 
Oxon, OX11 0QX, and University of Sussex, Physics and Astronomy Department, 
Brighton BN1 9QH, UK}
\newcommand{\uta}{Department of Physics, University of Texas at Austin,
Austin, TX 78712-0264}
\newcommand{\iusb}{Department of Physics and Astronomy, Indiana University, South Bend, IN}

\affiliation{	\ubc	}
\affiliation{	\bnl	}
\affiliation{	\carleton	}
\affiliation{	\uog	}
\affiliation{	\lu	}
\affiliation{	\lbnl	}
\affiliation{	\lanl	}
\affiliation{	\oxford	}
\affiliation{	\penn	}
\affiliation{	\queens	}
\affiliation{	\ralsuss	}
\affiliation{	\triumf	}
\affiliation{	\uw	}

\author{	S.N.~Ahmed	}\affiliation{	\queens	}									
\author{	A.E.~Anthony	}\affiliation{	\uta	}									
\author{	E.W.~Beier	}\affiliation{	\penn	}									
\author{	A.~Bellerive	}\affiliation{	\carleton	}									
\author{	S.D.~Biller	}\affiliation{	\oxford	}									
\author{	J.~Boger	}\affiliation{	\bnl	}									
\author{	M.G.~Boulay	}\affiliation{	\lanl	}									
\author{	M.G.~Bowler	}\affiliation{	\oxford	}									
\author{	T.J.~Bowles	}\affiliation{	\lanl	}									
\author{	S.J.~Brice	}\affiliation{	\lanl	}									
% ...
			
\collaboration{SNO Collaboration}
\noaffiliation
\date{\today}
\begin{abstract}
The Sudbury Neutrino Observatory (SNO) has precisely determined the total active
($\nu_{x}$) $^{8}$B solar neutrino flux without assumptions about the energy dependence of the 
$\nu_{e}$ survival probability.  The measurements were made with dissolved NaCl in the heavy water
to enhance the sensitivity and signature for neutral-current interactions.  The flux is found to be $\snoncfluxunc\times10^{6}$~cm$^{-2}$s$^{-1}$, in agreement
with previous measurements and standard solar models.  A global analysis of
these and other solar and reactor neutrino results yields 
$\Delta m^{2} = 7.1^{+1.2}_{-0.6}\times10^{-5}$~eV$^2$ and
$\theta = 32.5^{+2.4}_{-2.3}$ degrees.  Maximal mixing is rejected at the equivalent of \nsigmamax~standard deviations.
\end{abstract}
\maketitle
The Sudbury Neutrino Observatory (SNO)~\cite{sno_nim} detects $^8$B solar neutrinos
through the reactions
\begin{displaymath}
\begin{array}{l l l l}

\nu_e + d          & \rightarrow &
        p~ \!+ p~+ e^- & \qquad\text{(CC)}, \\
\nu_x + d          & \rightarrow &
        p + n + \nu_x        & \qquad\text{(NC)}, \\
\nu_x + e^-\!\!\!\!& \rightarrow &
        \nu_x + e^-                 & \qquad\text{(ES).}
\end{array}
\end{displaymath}
Only electron neutrinos produce charged-current interactions (CC),
while the neutral-current (NC) and elastic scattering (ES) reactions
have sensitivity to non-electron flavors.  The NC reaction measures
the total flux of all active neutrino flavors above a threshold of 2.2 MeV.

SNO previously measured the NC rate by observing neutron captures on deuterons, and found
that a Standard-Model description with an undistorted ${}^{8}$B neutrino spectrum
and CC, ES, and NC rates due solely to $\nu_{e}$ interactions
was rejected~\cite{ccprl,ncprl}.    This Letter presents
measurements of the CC, NC, and ES rates from SNO's dissolved salt
phase.

The addition of 2 tonnes of NaCl to the kilotonne of heavy water increased the neutron capture efficiency and the associated Cherenkov light.  
The solution was thoroughly mixed and a conductivity scan along the vertical
axis showed the NaCl concentration to be uniform within 0.5\%.

The data presented here were recorded between July 26, 2001 and October 10, 2002, totaling 254.2 live days.
 The number of raw triggers was 435,721,068 and the data set was  reduced to 3055 events after data reduction
similar to that in~\cite{ncprl} and analysis selection requirements.
Cherenkov event backgrounds from $\beta-\gamma$ decays were reduced with
an effective electron kinetic energy threshold $T_{\rm eff}$ $\geq$ 5.5 MeV and a fiducial volume
with radius $R_{\rm{fit}} \leq 550$ cm.
 
The neutron detection efficiency and response were calibrated with a ${}^{252}$Cf neutron source. The  neutron capture efficiency is shown in Fig.~1(a). The detection 
efficiency for NC reactions in the heavy water was $0.399 \pm 0.010~\mbox{(calibration)}\pm0.009~\mbox{(fiducial volume)}$ for $T_{\rm{eff}}$$\geq5.5$~MeV and $R_{\rm{fit}}$$\leq$550 cm, an increase of approximately a factor
of three from the pure D$_2$O phase. 
Calibration of the detector's optical and energy response has been updated to include time variation of the water transparency measurements made at various wavelengths throughout the running period.  A  normalization for photon detection efficiency based on ${}^{16}$N calibration data~\cite{n16_nim} and Monte Carlo calculations was used to set the absolute energy scale.  ${}^{16}$N data taken throughout the running period verified the gain drift (approximately 2\% per year) predicted by Monte Carlo calculations based on the optical measurements.
The energy response for electrons is characterized by a Gaussian function with resolution $\sigma_{T} = -0.145 + 0.392\sqrt{{T}_{e}}+0.0353 {T}_{e}$, where ${T}_{e}$ is the electron kinetic energy in MeV.  The energy scale uncertainty is $1.1\%$.

Neutron capture on $^{35}$Cl typically produces multiple $\gamma$ rays while the CC and ES reactions produce single electrons.  The greater isotropy of
the Cherenkov light from neutron capture events relative to CC and ES events allows good statistical separation of the event types.
This separation allows a precise measurement of the NC flux to be made independent of assumptions about the CC and ES energy spectra. 

\begin{figure}
\begin{center}
\includegraphics[width=3.73in]{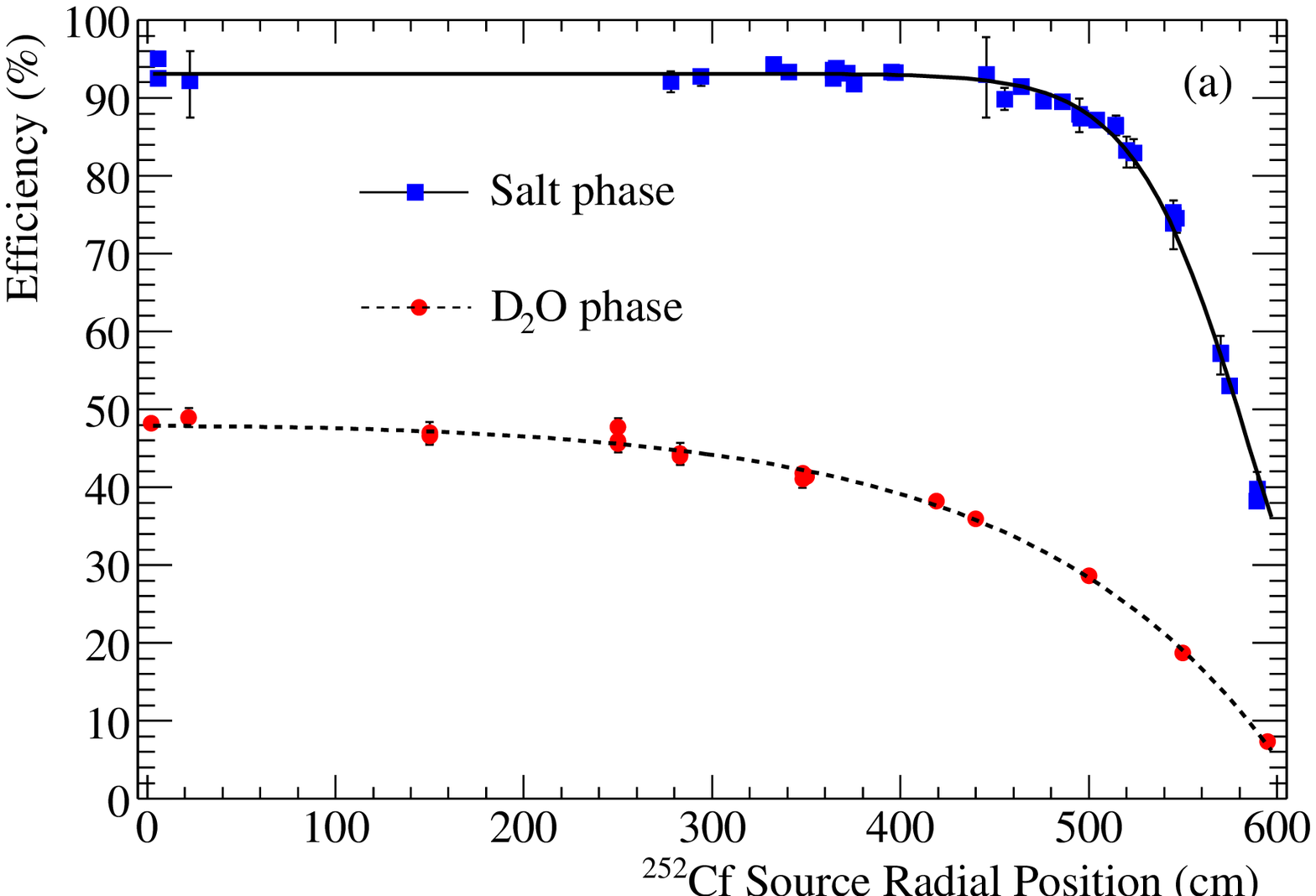}
\includegraphics[width=3.73in]{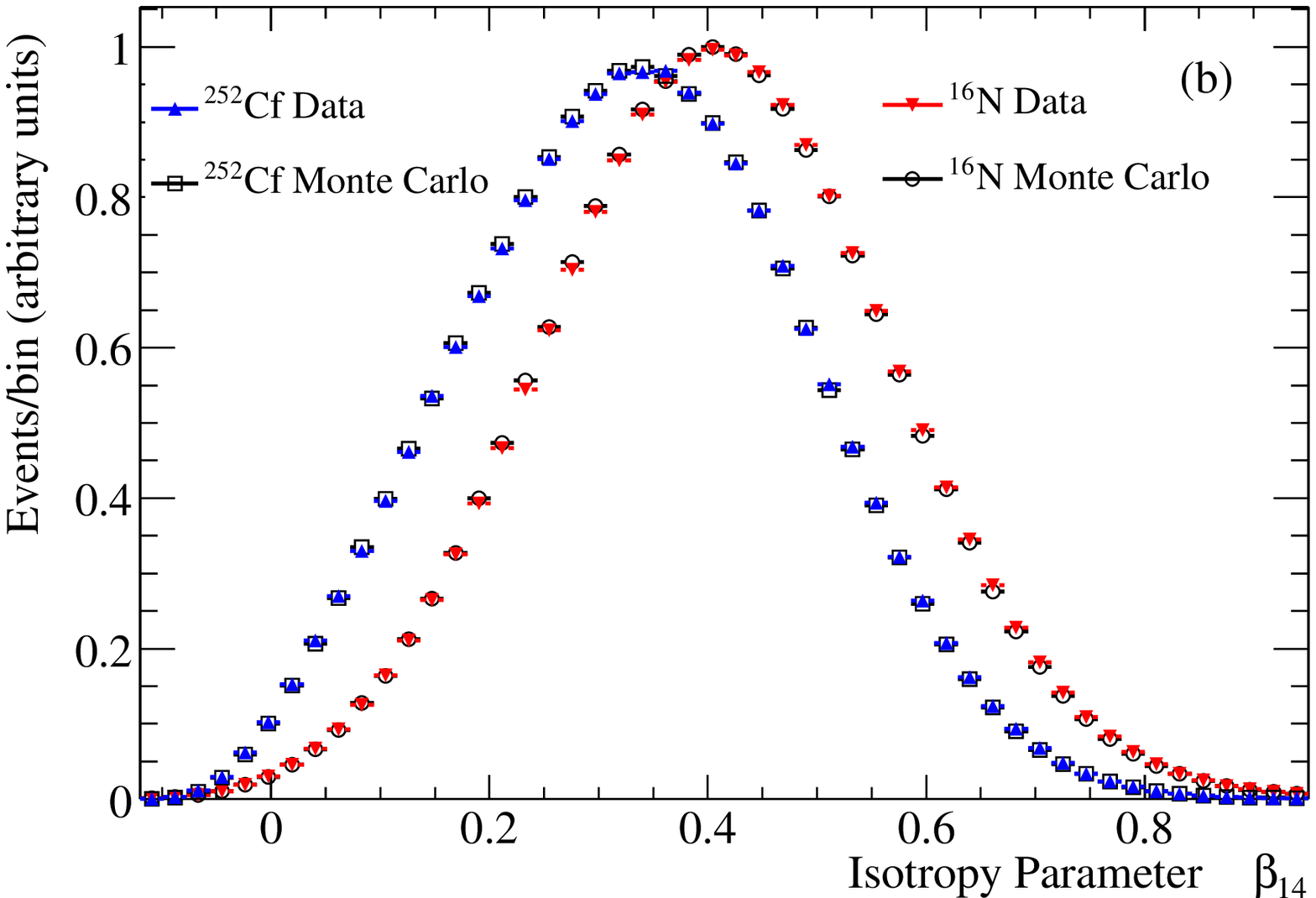}
\caption{\label{neutron_cal}(a) Neutron capture efficiency versus source radial position for the pure D$_2$O phase (capture on D) and salt phase (capture on Cl or D) deduced from a ${}^{252}$Cf source, with fits to an analytic function (salt) and to a neutron diffusion model (D$_2$O). (b) Event isotropy from data and Monte Carlo calculations of a ${}^{252}$Cf source and an ${}^{16}$N $\gamma$-ray source.}
\end{center}
\vspace{-4ex}
\end{figure}

The degree of the Cherenkov light isotropy is reflected in the pattern of photomultiplier-tube (PMT) hits.  Event isotropy was characterized by parameters $\beta_{l}$, the average value
 of the Legendre polynomial $P_{l}$ of the cosine of the angle between PMT hits~\cite{b14_footnote}.  The combination $\beta_{1}+4\beta_{4} \equiv \beta_{14}$ was selected as the measure of event isotropy to optimize the separation of NC and CC events.  Systematic uncertainty on $\beta_{14}$ distributions generated by Monte Carlo for 
signal events was evaluated by comparing ${}^{16}$N calibration data to Monte Carlo calculations~\cite{mott_footnote} for events
throughout the fiducial volume and running period. The uncertainty on the mean value of $\beta_{14}$ is $0.87\%$.  Comparisons of $\beta_{14}$ distributions from ${}^{16}$N events and neutron events from ${}^{252}$Cf to 
Monte Carlo calculations are shown in Fig.~\ref{neutron_cal}(b).  The Monte Carlo calculations of $\beta_{14}$ have also been verified with 19.8-MeV $\gamma$-ray events from a $^3$H($p$,$\gamma$)$^4$He source \cite{pt_nim}, high-energy electron events (dominated by CC and ES interactions) from the pure D$_2$O phase of the experiment, neutron events following muons, and with low-energy calibration sources.

Backgrounds are summarized in Table~\ref{letable}.
Low levels of the U and Th progeny ${}^{214}$Bi and ${}^{208}$Tl can create free neutrons from deuteron photodisintegration and low-energy
Cherenkov events from $\beta-\gamma$ decays.
{\textit{Ex-situ}} assays and {\textit{in-situ}} analysis techniques were employed to determine the average levels of uranium and thorium during the experiment~\cite{ncprl,mnox_nim,htio_nim,rn_nim}.   Results from these methods are consistent.  For the ${}^{232}$Th chain, the weighted mean of the {\textit{ex-situ}} MnO$_{\rm{x}}$ and the  {\textit{in-situ}}  measurement was used.  The ${}^{238}$U chain activity is dominated by Rn ingress, which is highly time-variable.  The {\textit{in-situ}} determination was used to estimate this background, because it provides the appropriate time weighting.  The average rate of background neutron production from activity in the D$_2$O region is  $0.72 ^{+0.24}_{-0.23}$ neutrons per day.  

Backgrounds from atmospheric neutrino interactions and ${}^{238}$U fission were estimated with the aid of the code NUANCE~\cite{casper} and from event
multiplicities. 

${}^{24}$Na (which originates from neutron activation of ${}^{23}$Na) can also emit $\gamma$ rays which photodisintegrate the deuteron.  Residual activation after calibrations and ${}^{24}$Na production in the water circulation system and in the heavy water in the ``neck'' of the vessel were determined and are included in Table~\ref{letable}.
 SNO is slightly sensitive to solar CNO neutrinos from the electron-capture decay of $^{15}$O and ${}^{17}$F.

Neutrons and $\gamma$ rays produced at the acrylic vessel and in the light water can propagate into the fiducial volume.  Radon progeny deposited on the surfaces of the vessel during construction can initiate ($\alpha$,$n$) reactions on  $^{13}$C, $^{17}$O, and $^{18}$O, and external $\gamma$ rays can photodisintegrate deuterium.  The enhanced neutron capture efficiency of salt makes these external-source neutrons readily apparent, and an additional distribution function was included in the analysis to extract this component (Table~\ref{letable}).   Previous measurements~\cite{ccprl,ncprl, snodn} with pure D$_2$O were less sensitive to this background source, and a preliminary evaluation indicates the number of these background events was within the systematic uncertainties reported.

The backgrounds from Cherenkov events inside and
outside the fiducial volume were estimated using calibration source data, measured activity, 
Monte Carlo calculations, and controlled injections of Rn into the detector.  These backgrounds were nearly negligible above the analysis energy
threshold and within the fiducial volume, and are included as  an uncertainty on the flux measurements.

A class of background events identified and removed from the analysis in the pure D$_2$O phase (``AV events'') reconstruct
near the acrylic vessel and were characterized by a nearly isotropic light distribution.  Analyses
of the pure D$_2$O and salt data sets limit this background to 5.4 events (68\% CL) for the present data.

\begingroup
\begin{table}
\squeezetable
\caption{\label{letable}Background events. The internal neutron and $\gamma$-ray backgrounds are constrained in the analysis. The external-source neutrons are reported from the fit.  The last two Table entries are included in the systematic uncertainty estimates.}
\begin{ruledtabular}
\begin{tabular}{ll}
Source                                  & Events                        \\ \hline
Deuteron photodisintegration              &  $73.1^{+24.0}_{-23.5}$               \\
${}^{2}$H($\alpha,\alpha$)$pn$            &  $2.8 \pm 0.7$                  \\
${}^{17,18}$O($\alpha$,$n$)                  &  $1.4 \pm 0.9$                       \\
Fission, atmospheric $\nu$ (NC +                    &                                    \\
\hspace{2em}sub-Cherenkov threshold CC)         &  $ 23.0 \pm 7.2$                   \\
Terrestrial and reactor $\bar{\nu}$'s   &  $2.3 \pm 0.8$                \\
Neutrons from rock                      &  $\leq 1$                       \\ 
${}^{24}$Na activation                  &  $8.4 \pm 2.3$                   \\
$n$ from CNO $\nu$'s                             &  $0.3 \pm 0.3$                   \\ \hline
Total internal neutron background       &  $111.3^{+25.3}_{-24.9}$               \\ \hline
Internal $\gamma$ (fission, atmospheric $\nu$) & $5.2 \pm 1.3$ \\  
${}^{16}$N decays                       & $< 2.5$ (68\% CL) \\ \hline
External-source neutrons (from fit)  &  $84.5^{+34.5}_{-33.6}$ \\ \hline
Cherenkov events from $\beta-\gamma$ decays   & $<14.7$  (68\% CL) \\ 
``AV events''                                          & $<5.4$ (68\% CL) \\ 
\end{tabular}
\end{ruledtabular}
\end{table}
\endgroup

\begin{figure}
\begin{center}
\psfrag{COSSUN}{~~~~~~$\cos \theta_{\odot}$}
\includegraphics[width=3.73in]{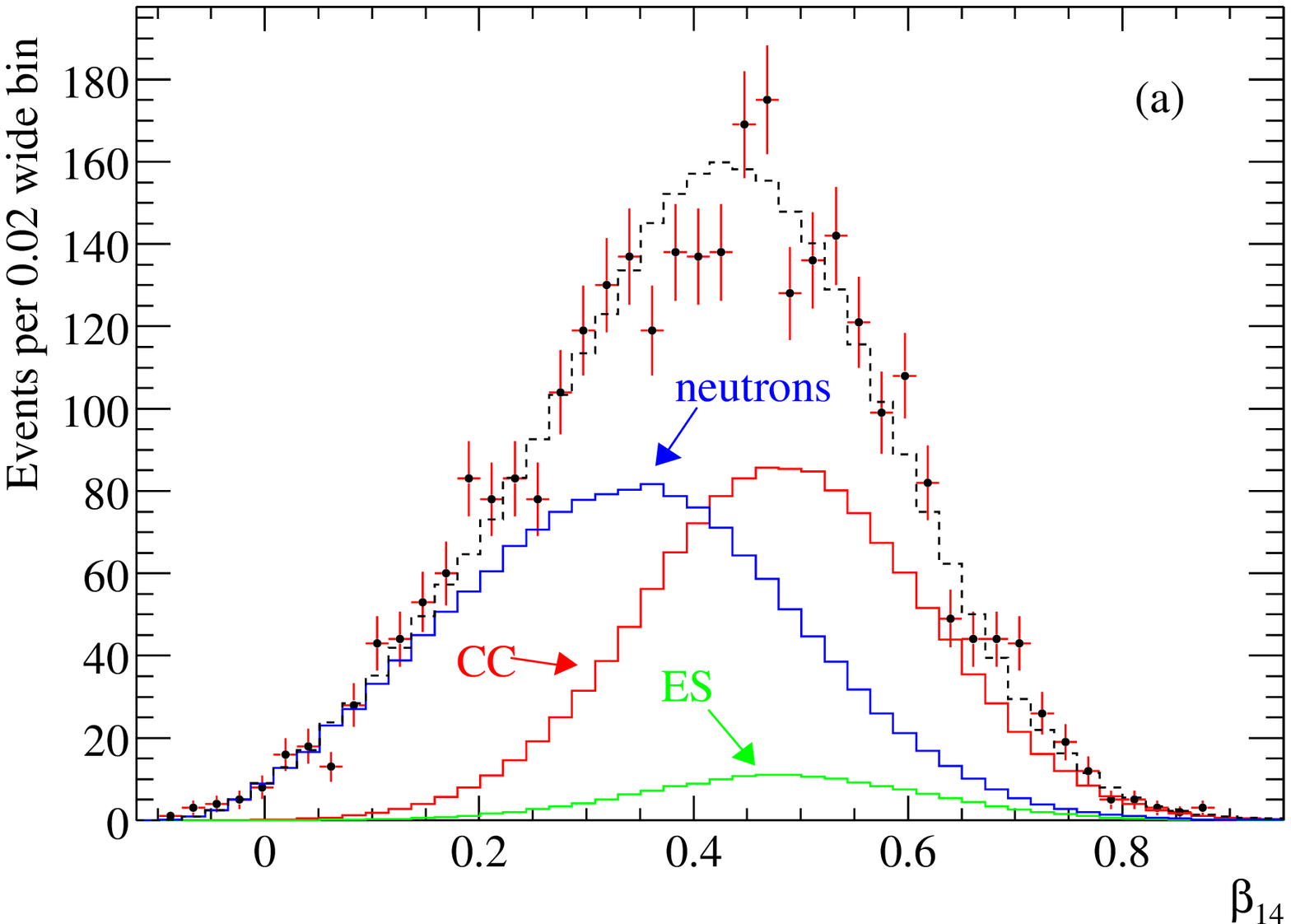}
\includegraphics[width=3.73in]{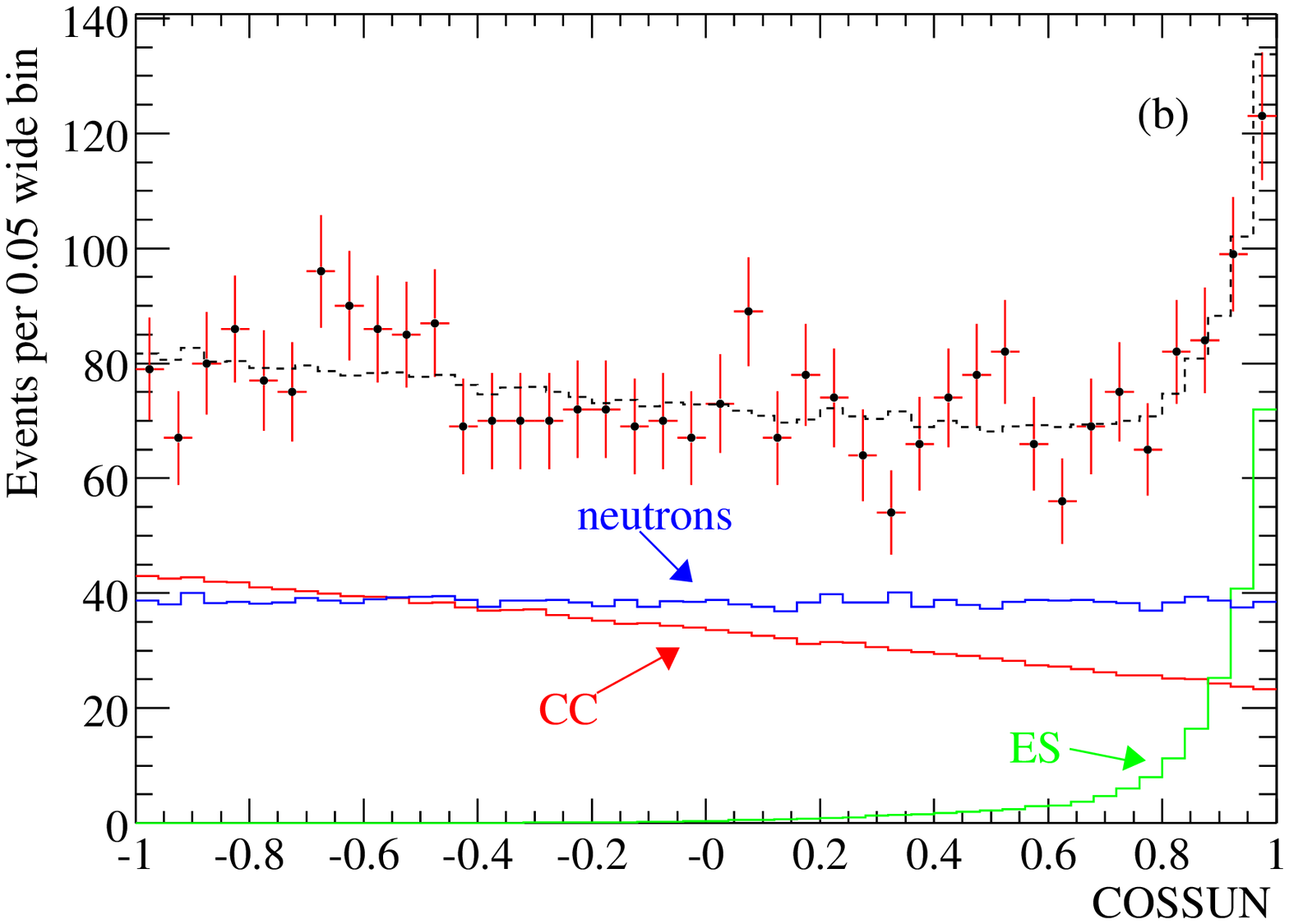}
\includegraphics[width=3.73in]{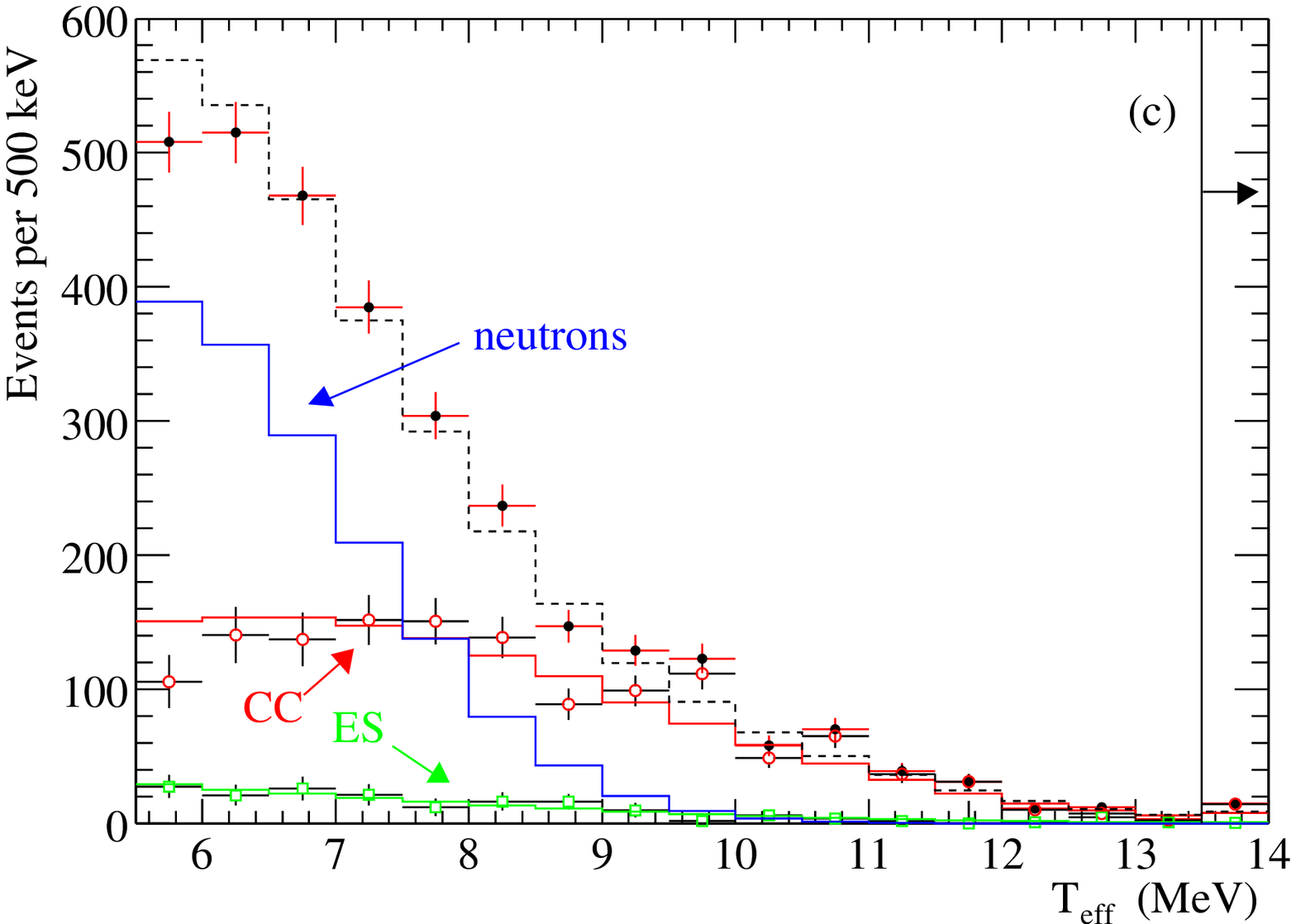}

\caption{\label{salt_data}Distribution of (a) $\beta_{14}$, (b) $\cos \theta_{\odot}$ and (c) kinetic energy, for the selected events.  The CC and ES spectra are extracted from the data using $\beta_{14}$ and $\cos \theta_{\odot}$ distributions in each energy bin.   Also shown are the Monte Carlo predictions for CC, ES, NC + internal and external-source neutron events, all scaled to the fit results.  The dashed lines represent the summed components.  All distributions are for events with $T_{\rm eff}$$\geq$5.5 MeV and R$_{\rm fit}$$\leq$ 550 cm. Differential systematics are not shown.}
\end{center}
\end{figure}

\begin{figure}
\begin{center}
\includegraphics[width=3.73in]{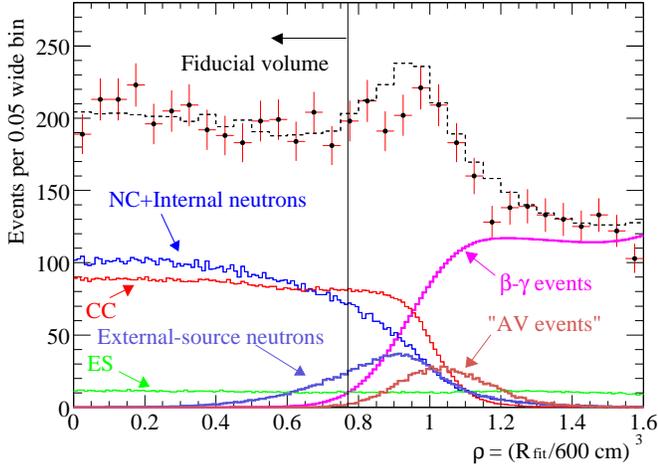}
\caption{\label{rho_comp}Volume-weighted radius  ($\rho$) distribution (acrylic vessel=1).  Shown are the distributions for CC, ES, NC+internal and external-source neutrons, scaled from the fit, and separately determined external background distributions.}
\end{center}
\vspace{-2ex}
\end{figure}

To minimize the possibility of introducing biases, a blind analysis procedure was used.  The data set used during the development of the analysis procedures
and the definition of parameters excluded an unknown fraction ($<30\%$) of the final data set, included an unknown admixture of muon-following neutron events, and included an unknown NC cross-section scaling factor. 
After fixing all analysis procedures and parameters, the blindness constraints were removed. The analysis was then performed on the `open' data set, statistically separating events into CC, NC, ES, and external-source neutrons using an extended maximum likelihood analysis based on the 
distributions of isotropy, cosine of the event direction relative to the vector from the Sun ($\cos\theta_{\odot}$), and radius within the detector.
This analysis differs from the analysis of the pure D$_2$O data~\cite{ccprl,ncprl,snodn} since the spectral distributions of the ES and CC events are not constrained
to the ${}^{8}$B shape, but are extracted from the data.  The extended maximum likelihood analysis  yielded $\nccfit$ CC, $\nesfit$ ES,
$\nncfit$ NC, and $\nncbefit$ external-source neutron events.  The systematic uncertainties on derived fluxes are shown in Table~\ref{errors}. 
The isotropy, $\cos\theta_{\odot}$, and kinetic energy distributions for the selected events are shown in Fig.~\ref{salt_data}, with statistical uncertainties only.
A complete spectral analysis including the treatment of differential systematic uncertainties will be presented in a future report.
The volume-weighted radial distributions [$\rho=(R_{\rm fit}/600~\rm{cm})^{3}$] are shown in Fig.~\ref{rho_comp}.

\begingroup
\squeezetable
\begin{table}
\caption{\label{errors}Systematic uncertainties on fluxes for the spectral shape unconstrained analysis of the salt data set. $\dagger$ denotes CC vs NC anti-correlation. }
\begin{ruledtabular}
\begin{tabular}{llllll}
  Source       & NC uncert. & CC uncert. & ES uncert. \\ 
         & (\%) & (\%) & (\%) \\  \hline 
    Energy scale   & -3.7,+3.6 & -1.0,+1.1 & $\pm1.8$ \\ 
Energy resolution   & $\pm1.2$ & $\pm0.1$ & $\pm0.3$ \\ 
Energy non-linearity   & $\pm0.0$ & -0.0,+0.1 & $\pm0.0$ \\ 
 Radial accuracy   & -3.0,+3.5 & -2.6,+2.5 & -2.6,+2.9 \\ 
Vertex resolution   & $\pm0.2$ & $\pm0.0$ & $\pm0.2$ \\ 
Angular resolution   & $\pm0.2$ & $\pm0.2$ & $\pm2.4$ \\ 
   Isotropy mean $\dagger$  & -3.4,+3.1 & -3.4,+2.6 & -0.9,+1.1 \\ 
Isotropy resolution   & $\pm0.6$ & $\pm0.4$ & $\pm0.2$ \\ 
Radial energy bias   & -2.4,+1.9 & $\pm0.7$ & -1.3,+1.2 \\ 
Vertex Z accuracy $\dagger$  & -0.2,+0.3 & $\pm0.1$ & $\pm0.1$ \\ 
Internal background neutrons   & -1.9,+1.8 & $\pm0.0$ & $\pm0.0$ \\ 
Internal background $\gamma$'s   & $\pm0.1$ & $\pm0.1$ & $\pm0.0$ \\ 
 Neutron capture   & -2.5,+2.7 & $\pm0.0$ & $\pm0.0$ \\ 
Cherenkov backgrounds   & -1.1,+0.0 & -1.1,+0.0 & $\pm0.0$ \\ 
``AV events''           & -0.4,+0.0& -0.4,+0.0 & $\pm0.0$ \\ \hline
Total experimental uncertainty &-7.3,+7.2 & -4.6,+3.8 &-4.3,+4.5 \\ \hline 
Cross section~\cite{crosssection} & $\pm 1.1$  & $\pm 1.2 $& $\pm 0.5$ 
\end{tabular}
\end{ruledtabular}
\end{table}
\endgroup

The fitted numbers of events give the equivalent $^8$B
fluxes~\cite{hep_footnote,es_crosssection}~(in units of $10^6~{\rm cm}^{-2} {\rm s}^{-1}$): 
\begin{eqnarray*}
\phi^{\text{SNO}}_{\text{CC}} & = & \snoccfluxunc \\
\phi^{\text{SNO}}_{\text{ES}} & = & \snoesfluxunc \\
\phi^{\text{SNO}}_{\text{NC}} & = & \snoncfluxunc~\mbox{,}
\end{eqnarray*} and the ratio of the ${}^{8}$B flux measured with
the CC and NC reactions is
\begin{equation*}
\frac{\phi^{\text{SNO}}_{\text{CC}}}{\phi^{\text{SNO}}_{\text{NC}}}  =  \snoccncratio. \\
\end{equation*}
 
Adding the constraint of an undistorted ${}^{8}$B energy spectrum to the analysis yields,
for comparison with earlier results~(in units of $10^6~{\rm cm}^{-2} {\rm s}^{-1}$): 
\begin{eqnarray*}
\phi^{\text{SNO}}_{\text{CC}} & = & \snoccfluxconc \\
\phi^{\text{SNO}}_{\text{ES}} & = & \snoesfluxconc \\
\phi^{\text{SNO}}_{\text{NC}} & = & \snoncfluxconc,
\end{eqnarray*}
\noindent consistent with the previous SNO measurements~\cite{ccprl,ncprl,constrained_sig}.
The difference between the CC results above (with and without constraint on the spectral shape) is $0.11\pm{0.05}~\mbox{(stat)}^{+0.06}_{-0.09}\mbox{(syst)}\times10^{6}$~cm$^{-2}$s$^{-1}$.

\begin{figure}
\begin{center}
\includegraphics[width=3.73in]{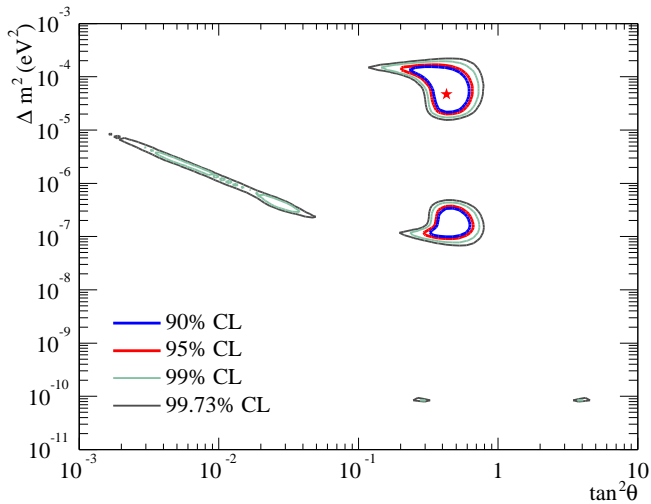}
\caption{\label{snoonlymsw}SNO-only neutrino oscillation contours, including pure D$_2$O day and night spectra, salt~CC, NC, ES fluxes, with ${}^{8}$B flux free and {\emph{hep}} flux fixed.
The best-fit point is $\Delta m^{2}=4.7\times10^{-5}$, $\tan^{2}\theta=0.43$, $f_{B}=1.03$, with $\chi^{2}$/d.o.f.=26.2/34. }
\end{center}
\end{figure}

\begin{figure}
\begin{center}
\includegraphics[width=3.73in]{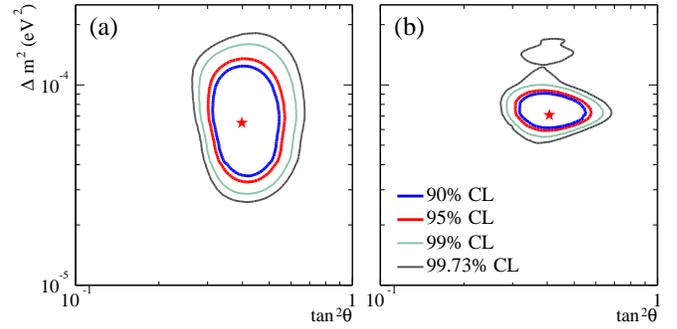}
\caption{\label{globalmsw}Global neutrino oscillation contours. (a) Solar global: D$_2$O day and night spectra, salt CC, NC, ES fluxes, SK, Cl, Ga.  The best-fit point is $\Delta m^2=6.5\times10^{-5}$, $\tan^{2}\theta=0.40$, $f_{B}=1.04$, with $\chi^{2}$/d.o.f.=70.2/81. (b) Solar global + KamLAND.  The best-fit point is $\Delta m^2=7.1\times10^{-5}$, $\tan^{2}\theta=0.41$, $f_{B} = 1.02$.  In both (a) and (b) the ${}^{8}$B flux is free and the {\em{hep}} flux is fixed. }
\end{center}
\end{figure}

The shape-unconstrained fluxes presented here combined with day and night
energy spectra from the pure D$_2$O phase~\cite{snodn} place constraints on
allowed neutrino flavor mixing parameters.  Two-flavor active neutrino
oscillation models predict the CC, NC and ES rates in SNO~\cite{mnspmsw}.
New analysis elements included Super-Kamiokande (SK) zenith spectra~\cite{superk}, updated Ga experiments~\cite{newsage,newgno}, improved
 treatment of $^8$B spectral systematic uncertainties, and
Cl and Ga cross-section correlations~\cite{fogli_d2o}.  
This improved oscillation analysis applied to the pure D$_2$O data reproduces other results~\cite{fogli_d2o,bahcall_d2o}.
A combined $\chi^2$ fit to SNO D$_2$O and salt data~\cite{snocompanion} alone yields the allowed regions in
$\Delta m^2$ and $\tan^2 \theta$ shown in Fig.~\ref{snoonlymsw}.  
In a global analysis of all solar neutrino data,
the allowed regions in parameter space shrink considerably and the LMA region is selected, as shown in
Fig.~\ref{globalmsw}(a).   A global analysis including the KamLAND reactor anti-neutrino results~\cite{kamland} shrinks the allowed region further, with a best-fit point of $\Delta m^{2} = 7.1^{+1.2}_{-0.6}\times10^{-5}$~eV$^2$ and
$\theta = 32.5^{+2.4}_{-2.3}$ degrees, where the errors reflect 1~$\sigma$ constraints on the 2-dimensional region~[Fig.~\ref{globalmsw}(b)].
With the new SNO measurements the allowed LMA region is constrained to only the lower band at $>99\%$ CL.
The best-fit point with one dimensional projection of the uncertainties in the individual parameters (marginalized uncertainties) is $\Delta m^{2} = 7.1^{+1.0}_{-0.3}\times10^{-5}$~eV$^2$ and $\theta = 32.5^{+1.7}_{-1.6}$ degrees. This disfavors maximal mixing at a level equivalent to \nsigmamax~$\sigma$.
In our analyses, the ratio $f_{B}$ of the total ${}^8$B flux to the SSM~\cite{bp2000} value was a free parameter,
while the total {\emph{hep}} flux was fixed at $9.3 \times 10^3$~cm$^{-2}$~s$^{-1}$.

In summary, we have precisely measured the total flux of active ${}^{8}$B  neutrinos 
from the Sun without assumptions about the energy dependence of the electron neutrino survival probability. 
The flux is in agreement with standard solar model calculations.  These results combined with global solar and reactor neutrino results reject the hypothesis of maximal
mixing at a confidence level equivalent to \nsigmamax~$\sigma$.

This research was supported by:  Canada: NSERC, Industry Canada, NRC,
Northern Ontario Heritage Fund, Inco, AECL, Ontario Power
Generation, HPCVL, CFI; US: Dept.\ of Energy; UK: PPARC. We thank the SNO technical
staff for their strong contributions. 
\vspace{6ex}

\bibliography{salt_prl}

\begin{thebibliography}{25}
\expandafter\ifx\csname natexlab\endcsname\relax\def\natexlab#1{#1}\fi
\expandafter\ifx\csname bibnamefont\endcsname\relax
  \def\bibnamefont#1{#1}\fi
\expandafter\ifx\csname bibfnamefont\endcsname\relax
  \def\bibfnamefont#1{#1}\fi
\expandafter\ifx\csname citenamefont\endcsname\relax
  \def\citenamefont#1{#1}\fi
\expandafter\ifx\csname url\endcsname\relax
  \def\url#1{\texttt{#1}}\fi
\expandafter\ifx\csname urlprefix\endcsname\relax\def\urlprefix{URL }\fi
\providecommand{\bibinfo}[2]{#2}
\providecommand{\eprint}[2][]{\url{#2}}

\bibitem[{\citenamefont{{SNO Collaboration}}(2000)}]{sno_nim}
\bibinfo{author}{\bibnamefont{{SNO Collaboration}}}, \bibinfo{journal}{Nucl.
  Instr. and Meth.} \textbf{\bibinfo{volume}{A449}}, \bibinfo{pages}{172}
  (\bibinfo{year}{2000}).

\bibitem[{\citenamefont{{SNO Collaboration}}(2001)}]{ccprl}
\bibinfo{author}{\bibnamefont{{SNO Collaboration}}}, \bibinfo{journal}{Phys.
  Rev. Lett.} \textbf{\bibinfo{volume}{87}}, \bibinfo{pages}{071301}
  (\bibinfo{year}{2001}).

\bibitem[{\citenamefont{{SNO Collaboration}}(2002{\natexlab{a}})}]{ncprl}
\bibinfo{author}{\bibnamefont{{SNO Collaboration}}}, \bibinfo{journal}{Phys.
  Rev. Lett.} \textbf{\bibinfo{volume}{89}}, \bibinfo{pages}{011301}
  (\bibinfo{year}{2002}{\natexlab{a}}).

\bibitem[{\citenamefont{{M.R. Dragowsky {\emph{et~al.}}}}(2002)}]{n16_nim}
\bibinfo{author}{\bibnamefont{{M.R. Dragowsky {\emph{et~al.}}}}},
  \bibinfo{journal}{Nucl. Instr. and Meth.} \textbf{\bibinfo{volume}{A481}},
  \bibinfo{pages}{284} (\bibinfo{year}{2002}).

\bibitem[{b14()}]{b14_footnote}
\bibinfo{note}{$\beta_l = \frac{2}{N(N-1)}\sum_{i=1}^{N-1} \sum_{j=i+1}^N
  P_l({\rm cos}\ \theta_{ij})$, with $\theta_{ij}$ the angle between hits $i$
  and $j$ from the event position and $N$ the number of hits.}

\bibitem[{mot()}]{mott_footnote}
\bibinfo{note}{A small correction to EGS4 was made to allow for the neglect of
  the Mott terms in the electron scattering cross-section and other
  approximations in the treatment of multiple scattering. We thank D.W.O.
  Rogers for providing EGSnrc simulations.}

\bibitem[{\citenamefont{{A.W.P. Poon {\emph{et~al.}}}}(2000)}]{pt_nim}
\bibinfo{author}{\bibnamefont{{A.W.P. Poon {\emph{et~al.}}}}},
  \bibinfo{journal}{Nucl. Instr. and Meth.} \textbf{\bibinfo{volume}{A452}},
  \bibinfo{pages}{115} (\bibinfo{year}{2000}).

\bibitem[{\citenamefont{{T.C.
  Andersen~{\emph{et~al.}}}}(2003{\natexlab{a}})}]{mnox_nim}
\bibinfo{author}{\bibnamefont{{T.C. Andersen~{\emph{et~al.}}}}},
  \bibinfo{journal}{Nucl. Instr. and Meth.} \textbf{\bibinfo{volume}{A501}},
  \bibinfo{pages}{399} (\bibinfo{year}{2003}{\natexlab{a}}).

\bibitem[{\citenamefont{{T.C.
  Andersen~{\emph{et~al.}}}}(2003{\natexlab{b}})}]{htio_nim}
\bibinfo{author}{\bibnamefont{{T.C. Andersen~{\emph{et~al.}}}}},
  \bibinfo{journal}{Nucl. Instr. and Meth.} \textbf{\bibinfo{volume}{A501}},
  \bibinfo{pages}{386} (\bibinfo{year}{2003}{\natexlab{b}}).

\bibitem[{\citenamefont{{I. Blevis~{\emph{et~al.}}}}()}]{rn_nim}
\bibinfo{author}{\bibnamefont{{I. Blevis~{\emph{et~al.}}}}},
  \bibinfo{note}{arXiv:nucl-ex/0305022, submitted to Nucl. Instr. and Meth..}

\bibitem[{\citenamefont{Casper}(2002)}]{casper}
\bibinfo{author}{\bibfnamefont{D.}~\bibnamefont{Casper}},
  \bibinfo{journal}{Nucl. Phys. Proc. Suppl.} \textbf{\bibinfo{volume}{112}},
  \bibinfo{pages}{161} (\bibinfo{year}{2002}).

\bibitem[{\citenamefont{{SNO Collaboration}}(2002{\natexlab{b}})}]{snodn}
\bibinfo{author}{\bibnamefont{{SNO Collaboration}}}, \bibinfo{journal}{Phys.
  Rev. Lett.} \textbf{\bibinfo{volume}{89}}, \bibinfo{pages}{011302}
  (\bibinfo{year}{2002}{\natexlab{b}}).

\bibitem[{cro()}]{crosssection}
\bibinfo{note}{Cross section uncertainty includes: $g_{A}$ (0.5\%), theoretical
  cross section (1\% for CC and NC, 0.3\% for CC/NC; S.
  Nakumura~{\emph{et~al.}} Nucl.\ Phys.\ {\bf A707}, 561 (2002)) and radiative
  corrections (0.3\% for CC, 0.1\% for NC; A. Kurylov, M.J. Ramsey-Musolf, and
  P. Vogel, Phys. Rev. {\bf C65} 05501 (2002)).}

\bibitem[{hep()}]{hep_footnote}
\bibinfo{note}{\emph{hep} neutrinos could also be present in the measured
  fluxes; the SSM contribution would be 0.5\%.}

\bibitem[{es_()}]{es_crosssection}
\bibinfo{note}{Electron neutrino cross sections are used to calculate all
  fluxes. The ES cross section is from J.N. Bahcall~{\emph{et~al.}}, Phys. Rev.
  {\bf{D51}}, 6146 (1995).}

\bibitem[{con()}]{constrained_sig}
\bibinfo{note}{A hypothesis test similar to~\cite{ncprl} disfavors no flavor
  transformation at greater than the equivalent of 7 $\sigma$.}

\bibitem[{mns()}]{mnspmsw}
\bibinfo{note}{Z.\ Maki, N.\ Nakagawa, and S.\ Sakata, Prog.\ Theor.\ Phys.\
  {\bf{28}} 870 (1962); V.\ Gribov and B.\ Pontecorvo, Phys.\ Lett.\ {\bf{B28}}
  493 (1969); S.\ P.\ Mikheyev and A.\ Yu.\ Smirnov, Sov.\ J.\ Nucl.\ Phys.\
  {\bf{42}} 913 (1985); L.\ Wolfenstein, Phys.\ Rev.\ {\bf{D17}} 2369 (1978).}

\bibitem[{\citenamefont{{S. Fukuda {\emph{et~al.}}}}(2002)}]{superk}
\bibinfo{author}{\bibnamefont{{S. Fukuda {\emph{et~al.}}}}},
  \bibinfo{journal}{Phys. Lett.} \textbf{\bibinfo{volume}{B539}},
  \bibinfo{pages}{179} (\bibinfo{year}{2002}).

\bibitem[{new({\natexlab{a}})}]{newsage}
\bibinfo{note}{V. Gavrin, 4th International Workshop on Low Energy and Solar
  Neutrinos, Paris, May 19--21, 2003.}

\bibitem[{new({\natexlab{b}})}]{newgno}
\bibinfo{note}{T. Kirsten, {\it Progress in GNO}, XXth Int.\ Conf. on Neutrino
  Physics and Astrophysics, Munich, May 25--30, 2002; to be published in Nucl.\
  Phys.\ B Proc.\ Suppl.}

\bibitem[{\citenamefont{{G.L. Fogli~{\emph{et~al.}}}}(2002)}]{fogli_d2o}
\bibinfo{author}{\bibnamefont{{G.L. Fogli~{\emph{et~al.}}}}},
  \bibinfo{journal}{Phys. Rev.} \textbf{\bibinfo{volume}{D66}},
  \bibinfo{pages}{053010} (\bibinfo{year}{2002}).

\bibitem[{\citenamefont{{J.N. Bahcall~{\emph{et~al.}}}}(2003)}]{bahcall_d2o}
\bibinfo{author}{\bibnamefont{{J.N. Bahcall~{\emph{et~al.}}}}},
  \bibinfo{journal}{JHEP} \textbf{\bibinfo{volume}{02}}, \bibinfo{pages}{009}
  (\bibinfo{year}{2003}).

\bibitem[{sno()}]{snocompanion}
\bibinfo{note}{See the SNO website http://sno.phy.queensu.ca for details.}

\bibitem[{\citenamefont{{K. Eguchi {\emph{et~al.}}}}(2003)}]{kamland}
\bibinfo{author}{\bibnamefont{{K. Eguchi {\emph{et~al.}}}}},
  \bibinfo{journal}{Phys. Rev. Lett.} \textbf{\bibinfo{volume}{90}},
  \bibinfo{pages}{021802} (\bibinfo{year}{2003}).

\bibitem[{\citenamefont{Bahcall et~al.}(2001)\citenamefont{Bahcall,
  Pinsonneault, and Basu}}]{bp2000}
\bibinfo{author}{\bibfnamefont{J.~N.} \bibnamefont{Bahcall}},
  \bibinfo{author}{\bibfnamefont{M.}~\bibnamefont{Pinsonneault}},
  \bibnamefont{and} \bibinfo{author}{\bibfnamefont{S.}~\bibnamefont{Basu}},
  \bibinfo{journal}{Astrophys. J.} \textbf{\bibinfo{volume}{555}},
  \bibinfo{pages}{990} (\bibinfo{year}{2001}).

\end{thebibliography}
\end{document}